\documentclass[prd,twocolumn,showpacs,amsmath,amssymb]{revtex4}
\usepackage{graphicx}
\usepackage{dcolumn}
\usepackage{bm}

\newcommand{\figwidth}{0.95\columnwidth}

\newcommand{\MWT}{${\cal M}$WT}
\newcommand{\vWT}{$\nu$WT}
\newcommand{\mpt}{\mbox{$\slash\kern-5.5pt p_T$}}
\newcommand{\mpx}{\mbox{$\slash\kern-5.5pt p_x$}}
\newcommand{\mpy}{\mbox{$\slash\kern-5.5pt p_y$}}
\newcommand{\x}{\hphantom{0}}

\begin{document}

\hspace{5.2in}\mbox{FERMILAB-PUB-06-354-E}

\title{Measurement of the top quark mass in the dilepton channel} 

%
\author{                                                                      
V.M.~Abazov,$^{35}$                                                           
B.~Abbott,$^{75}$                                                             
M.~Abolins,$^{65}$                                                            
B.S.~Acharya,$^{28}$                                                          
M.~Adams,$^{51}$                                                              
T.~Adams,$^{49}$                                                              
M.~Agelou,$^{17}$                                                             
E.~Aguilo,$^{5}$                                                              
S.H.~Ahn,$^{30}$                                                              
M.~Ahsan,$^{59}$
% ...
\\                                                                            
\vskip 0.30cm                                                                 
\centerline{(D\O\ Collaboration)}                                             
\vskip 0.30cm                                                                 
}                                                                             
\affiliation{                                                                 
\centerline{$^{1}$Universidad de Buenos Aires, Buenos Aires, Argentina}       
\centerline{$^{2}$LAFEX, Centro Brasileiro de Pesquisas F{\'\i}sicas,         
                  Rio de Janeiro, Brazil}                                     
\centerline{$^{3}$Universidade do Estado do Rio de Janeiro,                   
                  Rio de Janeiro, Brazil}                                     
\centerline{$^{4}$Instituto de F\'{\i}sica Te\'orica, Universidade            
                  Estadual Paulista, S\~ao Paulo, Brazil}                     
\centerline{$^{5}$University of Alberta, Edmonton, Alberta, Canada,           
                  Simon Fraser University, Burnaby, British Columbia, Canada,}
\centerline{York University, Toronto, Ontario, Canada, and                    
                  McGill University, Montreal, Quebec, Canada}                
\centerline{$^{6}$University of Science and Technology of China, Hefei,       
                  People's Republic of China}                                 
\centerline{$^{7}$Universidad de los Andes, Bogot\'{a}, Colombia}             
\centerline{$^{8}$Center for Particle Physics, Charles University,            
                  Prague, Czech Republic}                                     
\centerline{$^{9}$Czech Technical University, Prague, Czech Republic}         
\centerline{$^{10}$Center for Particle Physics, Institute of Physics,         
                   Academy of Sciences of the Czech Republic,                 
                   Prague, Czech Republic}                                    
\centerline{$^{11}$Universidad San Francisco de Quito, Quito, Ecuador}        
\centerline{$^{12}$Laboratoire de Physique Corpusculaire, IN2P3-CNRS,         
                   Universit\'e Blaise Pascal, Clermont-Ferrand, France}      
\centerline{$^{13}$Laboratoire de Physique Subatomique et de Cosmologie,      
                   IN2P3-CNRS, Universite de Grenoble 1, Grenoble, France}    
\centerline{$^{14}$CPPM, IN2P3-CNRS, Universit\'e de la M\'editerran\'ee,     
                   Marseille, France}                                         
\centerline{$^{15}$IN2P3-CNRS, Laboratoire de l'Acc\'el\'erateur              
                   Lin\'eaire, Orsay, France}                                 
\centerline{$^{16}$LPNHE, IN2P3-CNRS, Universit\'es Paris VI and VII,         
                   Paris, France}                                             
\centerline{$^{17}$DAPNIA/Service de Physique des Particules, CEA, Saclay,    
                   France}                                                    
\centerline{$^{18}$IPHC, IN2P3-CNRS, Universit\'e Louis Pasteur, Strasbourg,  
                    France, and Universit\'e de Haute Alsace,                 
                    Mulhouse, France}                                         
\centerline{$^{19}$Institut de Physique Nucl\'eaire de Lyon, IN2P3-CNRS,      
                   Universit\'e Claude Bernard, Villeurbanne, France}         
\centerline{$^{20}$III. Physikalisches Institut A, RWTH Aachen,               
                   Aachen, Germany}                                           
\centerline{$^{21}$Physikalisches Institut, Universit{\"a}t Bonn,             
                   Bonn, Germany}                                             
\centerline{$^{22}$Physikalisches Institut, Universit{\"a}t Freiburg,         
                   Freiburg, Germany}                                         
\centerline{$^{23}$Institut f{\"u}r Physik, Universit{\"a}t Mainz,            
                   Mainz, Germany}                                            
\centerline{$^{24}$Ludwig-Maximilians-Universit{\"a}t M{\"u}nchen,            
                   M{\"u}nchen, Germany}                                      
\centerline{$^{25}$Fachbereich Physik, University of Wuppertal,               
                   Wuppertal, Germany}                                        
\centerline{$^{26}$Panjab University, Chandigarh, India}                      
\centerline{$^{27}$Delhi University, Delhi, India}                            
\centerline{$^{28}$Tata Institute of Fundamental Research, Mumbai, India}     
\centerline{$^{29}$University College Dublin, Dublin, Ireland}                
\centerline{$^{30}$Korea Detector Laboratory, Korea University,               
                   Seoul, Korea}                                              
\centerline{$^{31}$SungKyunKwan University, Suwon, Korea}                     
\centerline{$^{32}$CINVESTAV, Mexico City, Mexico}                            
\centerline{$^{33}$FOM-Institute NIKHEF and University of                     
                   Amsterdam/NIKHEF, Amsterdam, The Netherlands}              
\centerline{$^{34}$Radboud University Nijmegen/NIKHEF, Nijmegen, The          
                  Netherlands}                                                
\centerline{$^{35}$Joint Institute for Nuclear Research, Dubna, Russia}       
\centerline{$^{36}$Institute for Theoretical and Experimental Physics,        
                   Moscow, Russia}                                            
\centerline{$^{37}$Moscow State University, Moscow, Russia}                   
\centerline{$^{38}$Institute for High Energy Physics, Protvino, Russia}       
\centerline{$^{39}$Petersburg Nuclear Physics Institute,                      
                   St. Petersburg, Russia}                                    
\centerline{$^{40}$Lund University, Lund, Sweden, Royal Institute of          
                   Technology and Stockholm University, Stockholm,            
                   Sweden, and}                                               
\centerline{Uppsala University, Uppsala, Sweden}                              
\centerline{$^{41}$Physik Institut der Universit{\"a}t Z{\"u}rich,            
                   Z{\"u}rich, Switzerland}                                   
\centerline{$^{42}$Lancaster University, Lancaster, United Kingdom}           
\centerline{$^{43}$Imperial College, London, United Kingdom}                  
\centerline{$^{44}$University of Manchester, Manchester, United Kingdom}      
\centerline{$^{45}$University of Arizona, Tucson, Arizona 85721, USA}         
\centerline{$^{46}$Lawrence Berkeley National Laboratory and University of    
                   California, Berkeley, California 94720, USA}               
\centerline{$^{47}$California State University, Fresno, California 93740, USA}
\centerline{$^{48}$University of California, Riverside, California 92521, USA}
\centerline{$^{49}$Florida State University, Tallahassee, Florida 32306, USA} 
\centerline{$^{50}$Fermi National Accelerator Laboratory,                     
            Batavia, Illinois 60510, USA}                                     
\centerline{$^{51}$University of Illinois at Chicago,                         
            Chicago, Illinois 60607, USA}                                     
\centerline{$^{52}$Northern Illinois University, DeKalb, Illinois 60115, USA} 
\centerline{$^{53}$Northwestern University, Evanston, Illinois 60208, USA}    
\centerline{$^{54}$Indiana University, Bloomington, Indiana 47405, USA}       
\centerline{$^{55}$University of Notre Dame, Notre Dame, Indiana 46556, USA}  
\centerline{$^{56}$Purdue University Calumet, Hammond, Indiana 46323, USA}    
\centerline{$^{57}$Iowa State University, Ames, Iowa 50011, USA}              
\centerline{$^{58}$University of Kansas, Lawrence, Kansas 66045, USA}         
\centerline{$^{59}$Kansas State University, Manhattan, Kansas 66506, USA}     
\centerline{$^{60}$Louisiana Tech University, Ruston, Louisiana 71272, USA}   
\centerline{$^{61}$University of Maryland, College Park, Maryland 20742, USA} 
\centerline{$^{62}$Boston University, Boston, Massachusetts 02215, USA}       
\centerline{$^{63}$Northeastern University, Boston, Massachusetts 02115, USA} 
\centerline{$^{64}$University of Michigan, Ann Arbor, Michigan 48109, USA}    
\centerline{$^{65}$Michigan State University,                                 
            East Lansing, Michigan 48824, USA}                                
\centerline{$^{66}$University of Mississippi,                                 
            University, Mississippi 38677, USA}                               
\centerline{$^{67}$University of Nebraska, Lincoln, Nebraska 68588, USA}      
\centerline{$^{68}$Princeton University, Princeton, New Jersey 08544, USA}    
\centerline{$^{69}$State University of New York, Buffalo, New York 14260, USA}
\centerline{$^{70}$Columbia University, New York, New York 10027, USA}        
\centerline{$^{71}$University of Rochester, Rochester, New York 14627, USA}   
\centerline{$^{72}$State University of New York,                              
            Stony Brook, New York 11794, USA}                                 
\centerline{$^{73}$Brookhaven National Laboratory, Upton, New York 11973, USA}
\centerline{$^{74}$Langston University, Langston, Oklahoma 73050, USA}        
\centerline{$^{75}$University of Oklahoma, Norman, Oklahoma 73019, USA}       
\centerline{$^{76}$Oklahoma State University, Stillwater, Oklahoma 74078, USA}
\centerline{$^{77}$Brown University, Providence, Rhode Island 02912, USA}     
\centerline{$^{78}$University of Texas, Arlington, Texas 76019, USA}          
\centerline{$^{79}$Southern Methodist University, Dallas, Texas 75275, USA}   
\centerline{$^{80}$Rice University, Houston, Texas 77005, USA}                
\centerline{$^{81}$University of Virginia, Charlottesville,                   
            Virginia 22901, USA}                                              
\centerline{$^{82}$University of Washington, Seattle, Washington 98195, USA}  
}                                                                             

\date{May 29, 2007}
           
\begin{abstract}
\vspace*{0.5cm}
We present a measurement of the top quark mass in the dilepton channel based on approximately 370~pb$^{-1}$ of data collected by the D\O\ experiment during Run II of the Fermilab Tevatron collider. We employ two different methods to extract the top quark mass. We show that both methods yield consistent results using ensemble tests of events generated with the D\O\ Monte Carlo simulation. We combine the results from the two methods to obtain a top quark mass $m_t = 178.1\pm8.2$~GeV. The statistical uncertainty is 6.7~GeV and the systematic uncertainty is 4.8~GeV.
\end{abstract}

\maketitle

The top quark mass is an important parameter in standard model~\cite{sm} predictions. For example, loops involving top quarks provide the dominant radiative corrections to the value of the $W$ boson mass. Precise measurements of the $W$ boson and top quark masses provide a constraint on the Higgs boson mass~\cite{Wrad}. 

At the Tevatron, top and antitop quarks are predominantly pair-produced. Top quarks decay to a $W$ boson and a $b$ quark. If the $W$ bosons from the top and the antitop quarks both decay leptonically (to $e\nu$ or $\mu\nu$) the final state consists of two charged leptons, missing transverse momentum ($\mpt$) from the undetected neutrinos, and two jets from the fragmentation of the $b$ quarks. We call this the dilepton channel. It has a relatively small branching fraction ($\approx 5$\%) but very low backgrounds. The measurement of the top quark mass in the dilepton channel is statistically limited. It provides an independent measurement of the top quark mass that can be compared with measurements in other $t\overline{t}$ decay channels, and a consistency check on the $t\overline{t}$ hypothesis in the dilepton channel. 

The D\O\ detector is a multipurpose collider detector~\cite{run2det}. The central tracker employs silicon microstrips close to the beam and concentric cylinders covered with scintillating fibers in a 2~T axial magnetic field. The liquid-argon/uranium calorimeter is divided into a central section covering pseudorapidity $|\eta|\leq1.1$ and two endcap calorimeters extending coverage to $|\eta|\leq 4.2$~\cite{run1det}, where $\eta=-\ln[\tan(\theta/2)]$ and $\theta$ is the polar angle with respect to the proton beam direction. The muon spectrometer consists of a layer of tracking detectors and scintillation trigger counters between the calorimeter and 1.8~T toroidal iron magnets, followed by two similar layers outside the toroids.

We present here two measurements that were carried out independently by two groups of analyzers. Both groups chose to optimize their analyses in different ways, one using a relatively loose event selection, the other taking advantage of the low background in top-antitop samples selected using tagging of $b$-quark jets. In the end, we combine the results from both analyses taking into account the correlations between the results.

The event selection is based on the measurement of the cross section for $t\overline{t}$-production in the dilepton channel\cite{xsec} with a few modifications. The analyses use about 370 pb$^{-1}$ of data from $p\overline{p}$ collisions at $\sqrt{s}$=1.96 TeV collected with the D\O\ detector at the Fermilab Tevatron collider.

We select events with two oppositely charged, isolated leptons ($e$ or $\mu$) with transverse momentum $p_T>15$~GeV and at least two jets with $p_T>20$~GeV. Electron candidates are isolated clusters of energy in the electromagnetic section of the calorimeter that agree in their profile with that expected from electromagnetic showers, based on Monte Carlo simulations, and that are matched with a charged particle track reconstructed in the central tracker. Electrons must be either in the central calorimeter (pseudorapidity $|\eta|<1.1$) or in the forward calorimeter ($1.5<|\eta|<2.5$). Muons are reconstructed as tracks in the muon spectrometer with $|\eta|<2$, matched to a charged particle track in the central tracker. They must be isolated from other activity in the calorimeter and in the tracker. Jets are reconstructed with the improved legacy cone algorithm\cite{jetalgo} with cone size $\Delta R = \sqrt{\Delta\eta^2+\Delta\phi^2} = 0.5$ and are restricted to $|\eta|<2.5$. All jets were corrected using the standard D\O\ jet energy scale corrections\cite{jetscale}.

We distinguish $e\mu$, $ee$, and $\mu\mu$ events. For $e\mu$ events we require $H_T>122$~GeV, where $H_T$ is the scalar sum of the larger of the two lepton $p_T$ values and the $p_T$ values of the leading two jets.
For $ee$ events we require sphericity $S> 0.15$ and missing transverse momentum $\mpt>35$--$40$~GeV, depending on the dielectron invariant mass $m(ee)$, and we reject events with $80<m(ee)<100$~GeV to reduce the background from $Z\rightarrow ee$ decays. Sphericity is defined as 1.5 times the sum of the first two eigenvalues of the normalized momentum tensor calculated using all electrons, muons and jets in the event.

For $\mu\mu$ events we require inconsistency with the $Z\rightarrow\mu\mu$ hypothesis based on the $\chi^2$ of a kinematic fit. In some $Z\rightarrow \mu\mu$ events a muon momentum is significantly mismeasured. These events are not consistent kinematically with $Z$ decays and they are therefore not eliminated by the kinematic fit. The mismeasured muon momentum gives rise to $p_T$ imbalance in the muon direction. We therefore require $\mpt>35$~GeV if the azimuthal angle between the leading muon and the direction of $\mpt$, $\Delta\phi(\mbox{$\mpt$},\mu) < 175^\circ$. We tighten the $\mpt$ requirement to 85 GeV if the leading muon and the $\mpt$ are approximately collinear in the transverse direction. 

For our mass measurements we use the following samples of events. 
The ``$b$-tag'' sample consists of events that have at least one jet that contains a secondary vertex tag with transverse decay length significance $\Lambda_{xy}>7$~\cite{btag}. This sample has very low backgrounds. The ``no-tag'' sample consists of events that have no such secondary vertex tags. The 26 events in these two samples consist of 20 $e\mu$ events, 5 $ee$, and 1 $\mu\mu$ event. 

The ``tight'' sample does not use the $b$-tagging information. It contains all $ee$ and $\mu\mu$ events that are in either the $b$-tag or the no-tag samples. For $e\mu$ events the tight sample requires the more restrictive cuts $H_T>140$~GeV, $\mpt>25$~ GeV and tighter electron identification cuts to reduce backgrounds. 
To increase the acceptance for dilepton decays, we also analyze a sample of events that requires only one well-identified lepton ($e$ or $\mu$) with $p_T>15$~GeV and an isolated track with $p_T>15$~GeV instead of the second identified lepton. The events must also have at least one jet with a secondary vertex tag, and $\mpt>15$--$35$~GeV, depending on lepton flavor and the invariant mass of the lepton+track system. We call this the ``$\ell$+track'' sample. Events with two well-identified leptons are vetoed from this sample so that there is no overlap between the $\ell$+track sample and the other dilepton samples. There are 6 $e$+track events and 3 $\mu$+track events in this sample. The observed event yields for each of the data samples are listed in Table \ref{tab:nem}.

\begin{table}[htp]
\caption{\label{tab:nem} Expected and observed dilepton event yields for $t\overline{t}$ production with $m_t=175$~GeV and the backgrounds from $WW$ and $Z$ production based on Monte Carlo, and from misidentified leptons (mis-id) based on collider data.} 
\begin{ruledtabular}
\begin{tabular}{lcccccccc}
Sample & $t\overline{t}$ & $WW$ & $Z$ & Mis-id & Total & Data \\ \hline
$\ell\ell$ no-tag  & \x$7.2$ & $1.1$\x & $2.6$\x & $2.2$ & $13.2\binom{+2.8}{-2.1}$ & 12 \\
$\ell\ell$ $b$-tag & \x$9.9$ & $0.05$ & $0.12$ & $0.09$ & $10.1\pm0.9$ & 14 \\
$\ell\ell$ tight   & $15.8$ & $1.1$\x & $2.4$\x & $0.5$ & $19.8\pm0.6$ & 21 \\
$\ell$+track & \x$6.3$ & $0.01$ & $1.8$\x & $0.4$\x & \x$8.5\pm0.3$ & \x9\\
\end{tabular}
\end{ruledtabular}
\end{table}

Monte Carlo samples are generated for nineteen values of the top quark mass between 120 and 230~GeV. The simulation uses {\sc alpgen}~\cite{alpgen} with {\sc cteq5l} parton distribution functions~\cite{CTEQ5} as the event generator, {\sc pythia}~\cite{pythia} for fragmentation and decay, and {\sc geant}~\cite{geant} for the detector simulation. No parton-shower matching algorithm was used in the generation of these event samples. We simulate diboson production with {\sc alpgen} and {\sc pythia} and $Z/\gamma^*\rightarrow\tau\tau$ processes with {\sc pythia}. The number of expected events are determined by applying the selection cuts to these Monte Carlo event samples. These samples are corrected for lepton, jet and $b$-tagging efficiencies determined from collider data.

The tagging efficiency for $b$-jets is measured in a data sample enhanced 
in heavy flavor jets by requiring at least one jet with a muon in each event. 
Monte-Carlo based corrections are applied to correct for sample biases. The
probability to tag a light-flavor jet is measured from collider data using 
events with a secondary vertex with negative decay length, meaning that the 
tracks forming the secondary vertex meet in the hemisphere that is on the opposite side of the primary vertex from the jet.

The energy of Monte Carlo jets is increased by 3.4\% in addition to the nominal jet energy scale corrections. This factor was determined by fitting the top mass and the jet energy scale in lepton+jets events and brings the invariant mass distribution of the two jets from the $W$ boson decay in lepton+jets Monte Carlo events in agreement with that observed in the data. 

Event yield normalizations for $Z\to ee$ and $Z\to \mu\mu$ are obtained from data. The number of events with misidentified leptons is dominated by jets misidentified as electrons. We construct a likelihood discriminant to distinguish electrons from misidentified jets based on the shape of the energy cluster in the calorimeter and the on the matched track. We determine the contamination by misidentified jets in our sample by fitting the distribution of this likelihood discriminant before we cut on it. Expected yields for signal and background are given in Table~\ref{tab:nem}.

We use only the two jets with the highest $p_T$ in this analysis. We assign these two jets to the $b$ and $\overline{b}$ quarks from the decay of the $t$ and $\overline{t}$ quarks. If we assume a value $m_t$ for the top quark mass, we can determine the pairs of $t$ and $\overline{t}$ momenta that are consistent with the observed lepton and jet momenta and $\mpt$. A solution refers to a pair of top-antitop quark momenta that is consistent with the observed event. For each assignment of observed momenta to the final state particles and for each hypothesized value of $m_t$, there may be up to four solutions. We assign a weight function $w(m_t)$ to each solution, as described below. Events for which no solution exists are rejected from our data and Monte Carlo event samples. The event yields in Table~\ref{tab:nem} include this additional selection requirement. Two events from the collider data are rejected with this requirement.

We consider each of the two possible assignments of the two jets to the $b$ and $\overline{b}$ quarks. We account for detector resolutions by repeating the weight calculation with input values for the lepton and jet momenta that are drawn from the detector resolution functions for objects with the observed momenta. We refer to this procedure as resolution sampling. For each event we obtain a weight $W(m_t)=1/N\times\sum_{j=1}^N\sum_{i=1}^n w(m_t)_{ij}$ by summing over all $n$ solutions and averaging over $N$ resolution samples. This weight characterizes the likelihood that the event is produced in the decay of a $t\overline{t}$ pair as a function of $m_t$. 

The techniques we use are similar to those used by the D\O\ Collaboration to measure the top quark mass in the dilepton channel using Run I data~\cite{RunI}. The data are analyzed using two different methods that differ in the event samples that they are based on, in the calculation of the event weight, and in the algorithm that compares the weights for the observed events to Monte Carlo predictions to extract the top quark mass.

The matrix-element weighting technique (\MWT) follows the ideas proposed by Dalitz and Goldstein~\cite{DalitzGoldstein} and Kondo~\cite{Kondo}. The solution weight is 
$$w(m_t)=f(x)f(\overline{x})p(E_\ell^*|m_t)p(E_{\overline{\ell}}^*|m_t),$$ 
where $f(x)$ is the parton distribution function of the proton and $x$ ($\overline{x}$) is the momentum fraction carried by the initial (anti)quark. The quantity $p(E_\ell^*|m_t)$ is the probability that the lepton has energy $E_\ell^*$ in the top quark rest frame for the hypothesized top quark mass $m_t$.

For each event we use the value of the hypothesized top quark mass $m_{\rm peak}$ at which $W(m_t)$ reaches its maximum as the estimator for the mass of the top quark. We  generate probability density functions of $m_{\rm peak}$ for a range of top quark masses using Monte Carlo simulations. We call these distributions templates. 
To compute the contribution of backgrounds to the templates, we use $Z\rightarrow\tau\tau$ and $WW$ Monte Carlo events. Backgrounds arising from detector signals that are misidentified as electrons or muons are estimated from collider data samples. 

We compare the distribution of $m_{\rm peak}$ for the observed events to these templates using a binned maximum likelihood fit. The likelihood is calculated as 
$$L(m_t)=\prod_{i=1}^{n_{\rm bin}}\left[\frac{n_ss_i(m_t)+n_bb_i}{n_s+n_b}\right]^{n_i},$$
where $n_i$ is the number of data events observed in bin $i$, $s_i(m_t)$ is the normalized signal template contents for bin $i$ at top quark mass $m_t$, $b_i$ is the normalized background template contents for bin $i$. The product runs over all $n_{\rm bin}$ bins. The background template consists of events from all background sources added in the expected relative proportions. The signal-to-background fraction is fixed to $n_s/n_b$ with the numbers of signal and background events ($n_s$, $n_b$) taken from Table \ref{tab:nem}.

To calibrate the performance of our method, we generate a large number of simulated experiments for several input top quark mass values. We refer to each of these experiments as an ensemble. Each ensemble consists of as many events of each type as we have in our collider data sample. A given event is taken from the signal and background samples with probabilities that correspond to the fraction of events expected from each sample. We use a quadratic function of $m_t$ to fit the $-\ln L$ points to thirteen mass points centered on the point with the smallest value of $-\ln L$. The distribution of measured top quark mass values from the ensemble fits gives an estimate of the parent distribution of our measurement. The ensemble test results indicate that the measured mass tracks the input mass with an offset of $1.9\pm0.8$~GeV, which we correct for in the final result. 

In general, the tails of the likelihood distribution for an ensemble are not well approximated by a Gaussian. Thus it is necessary to restrict the range of mass points that is included in the fit to points near the observed minimum in $-\ln L$. For small data samples, however, there is a substantial statistical uncertainty in the computed likelihood values which can be reduced by increasing the number of mass points used in the fit. Thus the range of mass points that are included in the likelihood fit must be optimized for the observed data sample size to obtain the best possible agreement between measured top quark mass and input top quark mass. This was done for both analyses based on Monte Carlo ensembles that contain exactly as many events as we observe in the data.

\begin{figure}[htp]
\includegraphics[width=\figwidth]{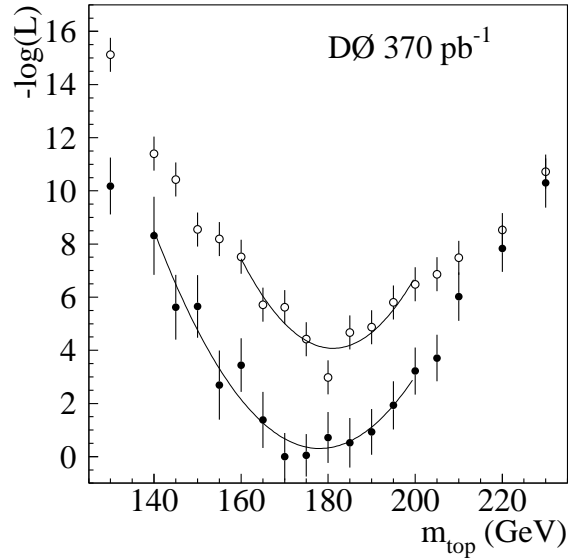}
\caption{\label{fig:fit} Joint likelihoods from the \MWT\ analysis (closed circles) and the \vWT\ analysis (open circles). The minima of the likelihood curves do not include the correction for the offset in the response.}
\end{figure}

\begin{figure}[htp]
\includegraphics[width=\figwidth]{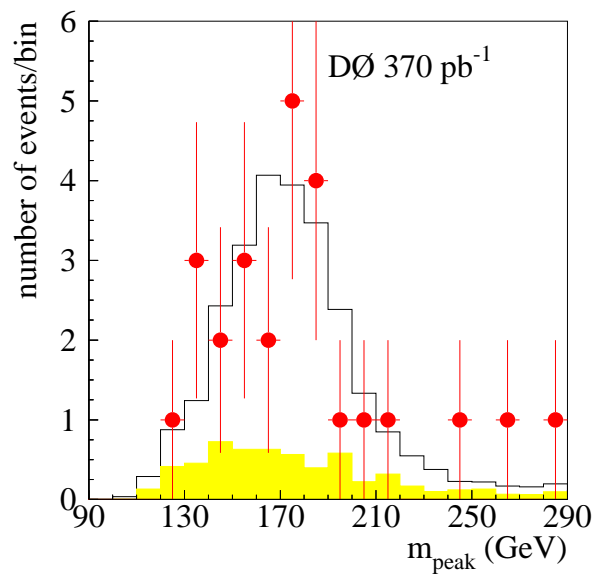}
\caption{\label{fig:peak} Distribution of $m_{\rm peak}$ from the \MWT\ analysis (circles) compared to the sum of Monte Carlo templates for the no-tag and $b$-tag channels and all lepton flavors for $m_t=180$~GeV (open histogram). The shaded histogram indicates the background contribution.}
\end{figure}

The \MWT\ analysis uses the no-tag and $b$-tag samples of events. Separating out the very-low-background $b$-tagged events improves the precision of the result. 
The analysis is performed with separate templates for $ee$, $e\mu$, and $\mu\mu$ events and separate signal-to-background fractions for events without a $b$-tag and $\geq1$ $b$-tags. The maximum of the joint likelihood for all events, shown in Fig.~\ref{fig:fit}, corresponds to $m_t=176.2\pm9.2(\mbox{stat})$~GeV after the offset correction. Figure \ref{fig:peak} shows the distribution of $m_{\rm peak}$ from collider data compared to the sum of Monte Carlo templates with $m_t=180$~GeV. 

The neutrino weighting technique (\vWT) ignores the measured $\mpt$ in reconstructing the event. Instead we assume a representative range of values for the pseudorapidities of the two neutrinos and the solution weight 
$$w(m_t)=\frac{1}{N_\eta} \sum_{i=1}^{N_\eta} \exp\left[\frac{-({\mpx}_i-\mpx)^2}{2\sigma_x^2}\right] \exp\left[\frac{-({\mpy}_i-\mpy)^2}{2\sigma_y^2}\right]$$ 
characterizes the consistency of the resulting solutions with the observed $\mpt$. The sum is over the $N_{\eta}$ steps of neutrino rapidity values, ${\mpx}_i$ and ${\mpy}_i$ are the $x$ and $y$ components of the sum of the neutrino momenta computed for step $i$, and $\sigma_x$ and $\sigma_y$ are the measurement resolutions for ${\mpx}$ and ${\mpy}$. We then normalize the event weight $W(m_t)$ over the range $80<m_t<330$~GeV and integrate it over ten bins in $m_t$. Every event is thus characterized by a 9-component vector $\overrightarrow{W}=(W_1,...,W_9)$ (the $10^{th}$ bin is fixed by the first nine and the normalization condition). We compare the vectors from the collider data events to sets of $N$ Monte Carlo events generated with different values of $m_t$ by computing the signal probability 
$$f_s(\overrightarrow{W}|m_t) = \frac{1}{N} \sum_{j=1}^N \prod_{i=1}^9 \frac{\exp[-(W_i-W_{ij}^{MC})^2/2h^2]}{\int_0^1 \exp[-(W'-W_{ij}^{MC})^2/2h^2]dW'},$$ 
where $\overrightarrow{W}_{j}^{MC}$ is the vector of weights from Monte Carlo event $j$. The value of the resolution parameter $h$ is optimized using ensemble tests based on simulated events to give the best agreement between input mass and measured mass. We compute a similar probability $f_b(\overrightarrow{W})$ for backgrounds and combine them in the likelihood 
\begin{eqnarray*}
L(m_t,\overline{n}_b,n)=G(n_b-\overline{n}_b,\sigma)P(n_s+n_b,n)\\
\times\prod_{i=1}^n\left[\frac{n_sf_s(\overrightarrow{W}_i|m_t)+n_bf_b(\overrightarrow{W}_i)}{n_s+n_b}\right],
\end{eqnarray*}
which we optimize with respect to $m_t$, the number of signal events $n_s$, and the number of background events $n_b$. $G$ is a gaussian constraint on the difference between $n_b$ and the expected number of background events $\overline{n}_b$, and $P$ is a Poisson constraint on $n_s+n_b$ to the number of events $n$ observed in data. 

The \vWT\ analysis uses the tight sample and the $\ell$+track sample. 
The analysis is performed with separate templates for $ee$, $e\mu$, and $\mu\mu$ events in the tight sample and the two lepton flavors in the $\ell$+track sample. We fit the $-\ln L$ points for values of $m_t$ within 20 GeV of the point with the smallest value of $-\ln L$ with a quadratic function of $m_t$. The performance of the \vWT\ algorithm is checked using ensemble tests as described for the \MWT\ algorithm. The average measured values of $m_t$ track the input values with an offset of $1.7\pm0.2$~GeV. For the \vWT\ analysis, the maximum of the joint likelihood of all events (Fig.~\ref{fig:fit}) corresponds to $m_t=179.5\pm7.4(\mbox{stat})$~GeV after the offset correction. 

We also use ensemble tests to study the size of systematic uncertainties (see Table~\ref{tab:summary}). By far the largest systematic uncertainty originates from the uncertainty in the calibration of the jet energy scale of 4.1\%. We determine the effect of the uncertainty on the measurement by generating ensemble tests with the jet energy scale increased and decreased by one standard deviation. 

We estimate the sensitivity of the result to uncertainties in the parton distribution functions by analyzing ensembles generated with a range of available parton distribution function sets.
The next to largest uncertainty originates from the modeling of gluon radiation in the Monte Carlo. Gluon radiation can give rise to additional jets in the event. In the data about one third of the events have more than two jets. The two analyses used different procedures to estimate this effect. For the \vWT\ analysis, events with three reconstructed jets from $t\bar{t}+1$ parton events generated with {\sc alpgen} were analyzed in ensemble tests with templates derived from $t\bar{t}$ events with only two jets and the difference in reconstructed top quark mass was applied as an uncertainty to the fraction of events with more than two jets. In the \MWT\ analysis the fraction of events with only two jets was varied in ensemble tests within the range that is consistent with the jet multiplicity spectrum observed in the data and analyzed with the nominal templates. The observed variation in the result was applied as systematic error.

We estimate the effect of uncertainties in the shape of the background distributions to determine the background uncertainty. For the \MWT\ analysis we also perform tests with ensembles in which we varied the background fraction, which was fixed in the mass fit, by its uncertainty.
For the $\ell$+track sample, the heavy flavor content in the background is a significant source of uncertainty. This only contributed to the \vWT\ analysis.
The finite size of the Monte Carlo samples limits the statistical precision with which we can extract the top quark mass. This is accounted for in the Monte Carlo statistics uncertainty.
Finally, we generated ensembles with varied jet and muon momentum resolutions to estimate the effect of their uncertainties. The resulting uncertainties for the \vWT\ analysis are quoted in Table~\ref{tab:summary}. The effect on the result of the \MWT\ analysis was negligible. 

We follow the method for combining correlated measurements from Ref.~\cite{Lyons} in combining the results from the \MWT\ and \vWT\ analyses. We determine the statistical correlation between the two measurements using ensemble tests. The correlation factor between the two analyses is $0.35$. The systematic uncertainties from each source in Table~\ref{tab:summary} are taken to be completely correlated between the two analyses. The results of the combination are also listed in Table~\ref{tab:summary}.

\begin{table}[ht]
\caption{\label{tab:summary}Summary of dilepton mass measurements.}
\begin{ruledtabular}
\begin{tabular}{lrrrrrr}
 & \MWT & \vWT & Combined \\ \hline
Top quark mass          & 176.2 & 179.5 & 178.1 GeV \\ \hline
Statistical uncertainty &   9.2 &   7.4 &   6.7 GeV \\
Systematic  uncertainty &   3.9 &   5.6 &   4.8 GeV \\ \hline
\ Jet energy scale      &   3.6 &   4.8 &   4.3 GeV \\
\ Parton distribution 
functions               &   0.9 &   0.7 &   0.8 GeV \\
\ Gluon radiation       &   0.8 &   2.0 &   1.5 GeV \\
\ Background            &   0.2 &   1.4 &   0.9 GeV \\ 
\ Heavy flavor content  &   --- &   0.6 &   0.3 GeV \\
\ Monte Carlo statistics &   0.8 &   1.0 &   0.9 GeV \\ 
\ Jet resolution        &   --- &   0.6 &   0.3 GeV \\
\ Muon resolution       &   --- &   0.4 &   0.2 GeV \\ \hline
Total uncertainty       &  10.0 &   9.3 &   8.2 GeV \\
\end{tabular}
\end{ruledtabular}
\end{table}

In conclusion, we measure the top quark mass in the dilepton channel. We obtain $m_t=178.1\pm6.7(\mbox{stat})\pm4.8(\mbox{syst})$~GeV as our best estimate of 
the top quark mass. This is in good agreement with the world average $m_t=172.5\pm2.3$~GeV~\cite{world}, based on Run I and Run II data collected by the CDF and D\O\ Collaborations.

%
We thank the staffs at Fermilab and collaborating institutions, 
and acknowledge support from the 
DOE and NSF (USA);
CEA and CNRS/IN2P3 (France);
FASI, Rosatom and RFBR (Russia);
CAPES, CNPq, FAPERJ, FAPESP and FUNDUNESP (Brazil);
DAE and DST (India);
Colciencias (Colombia);
CONACyT (Mexico);
KRF and KOSEF (Korea);
CONICET and UBACyT (Argentina);
FOM (The Netherlands);
PPARC (United Kingdom);
MSMT (Czech Republic);
CRC Program, CFI, NSERC and WestGrid Project (Canada);
BMBF and DFG (Germany);
SFI (Ireland);
The Swedish Research Council (Sweden);
Research Corporation;
Alexander von Humboldt Foundation;
and the Marie Curie Program.

\end{document}